# Capillary-driven indentation of a microparticle into a soft, oil-coated substrate

Justin D. Glover and Jonathan T. Pham[*]

Small scale contact between a soft, liquid-coated layer and a stiff surface is common in many situations, from synovial fluid on articular cartilage to adhesives in humid environments. Moreover, many model studies on soft adhesive contacts are conducted with soft silicone elastomers, which possess uncrosslinked liquid molecules (i.e. silicone oil) when the modulus is low. We investigate how the thickness of a silicone oil layer on a soft substrate relates to the indentation depth of glass microspheres in contact with crosslinked PDMS, which have a modulus of <10 kPa. The particles indent into the underlying substrate more as a function of decreasing oil layer thickness. This is due to the presence of the liquid layer at the surface that causes capillary forces to push down on the particle. A simple model that balances the capillary force of the oil layer and the particle-substrate adhesion with the elastic and surface tension forces from the substrate is proposed to predict the particle indentation depth.

## Introduction

Small scale contact with a soft, liquid-coated surface is common in many natural and industrial processes. In many cases, the presence of the liquid layer is critical for the system to perform its function. For example, synovial fluid in joints helps to reduce friction of contacting articular cartilage.[1,2] In nature, insects often rely on small scale adhesion with liquid layers; an oily secretion from small structures on insect feet leads to capillary-enhanced adhesion.[3-7] This type of mechanism has been exploited for developing bioinspired adhesives.[8-10] The importance of liquid capillarity on small scales is also demonstrated in mechanical characterization methods like atomic force microscopy; in a humid environment, condensation around the tip causes a downward capillary force on the cantilever.[11,12] However, capillarity can come from a solid when the contact is small on a sufficiently soft substrate. Small and soft is defined by the elastocapillary length $L_{EC} = \Upsilon/E$, where $\Upsilon$ is the surface tension of the solid and $E$ is the Young's modulus. When the characteristic size scales are near $L_{EC}$, surface forces have a significant effect relative to elastic restoring forces.[13-26] Hence, it would be beneficial to investigate a situation that includes both liquid and solid capillarity.

Recently, there has been a growing interest in the role of solid capillary forces for small scale adhesion and contact of soft materials. From an experimental perspective, many studies on elastocapillary surface deformations are conducted with soft crosslinked silicones (e.g. polydimethylsiloxane, PDMS). These materials often possess a significant fraction of uncrosslinked molecules (e.g. silicone oil), which can diffuse out of the network.[13, 27-30] This modifies the contact behavior by introducing liquid molecules. For example, these oil molecules are able to transfer from a PDMS surface to a contacting indenter, reducing the adhesion or friction between the surfaces.[28, 31] Near the elastocapillary scale, oil molecules have been reported to form a pure liquid zone near the contact line, allowing for lower deformations of the elastic network while accommodating the interfacial tensions.[13, 27, 32]

Although the interaction of a stiff microsphere with low surface tension silicone oil or with soft solid PDMS has become fairly well described, a mixed contact including both liquid and solid is less understood. When a glass microsphere is placed on silicone oil, the sphere is drawn into the liquid to lower the interfacial tension; that is, it becomes engulfed in the liquid and the interfacial tensions define the position of the particle. If the oil is transformed into a soft elastomer by crosslinking, a resistance to wetting the microparticle arises in the form of elasticity; this leads to a meniscus forming around the sphere instead of being fully cloaked. Adhesion between the microsphere and the crosslinked network promotes contact, whereas elastic restoring forces oppose it. Near $L_{EC}$, it has been shown that solid surface tension also resists indentation while pure oil zones can promote indentation. Capillary forces from an immiscible liquid have also been shown to increase adhesion between two soft solids.[33, 34] However, it is not clear how the amount of a low surface tension oil near the surface affects indentation of a small microparticle due to liquid capillarity.

Here we systematically investigate the indentation of glass microspheres placed on a low modulus elastomeric surface while controlling the amount of oil of the same composition. By varying the thickness of the oil layer, we find that the indentation depth depends on how thick the layer is relative to the particle size. Our results are fit reasonably well with an analytical model based on the elastic deformation and solid surface tension of the substrate, balanced by the capillary forces of the oil layer.

## Results and discussion

In our experiments, glass microspheres with a radius range of $R \approx 9\text{-}31$ μm are sprinkled onto a soft PDMS surface coated with a layer of silicone oil. We first prepare a soft PDMS substrate using Sylgard 184 at a base to crosslinker ratio of 60 to 1. This mixing ratio yields a Young's modulus on the order of a few kPa.[13, 29, 35, 36] The surface is prepared by spin-coating the uncured mixture on a glass coverslip to a thickness of $t_{solid} \sim 90$ μm (Fig. 1A and Fig. S1) and then cured; this is sufficiently thin to obtain high quality confocal images with an inverted microscope looking through the sample. On the other hand, since the values of the relative contact size and indentation depth $a_{solid}/t_{solid}$ and $\delta/t_{solid}$ are small, we expect this to be sufficiently thick to neglect the finite

a. Department of chemical and materials engineering, University of Kentucky, Lexington, KY 40506, USA
*Email: Jonathan.Pham@uky.edu

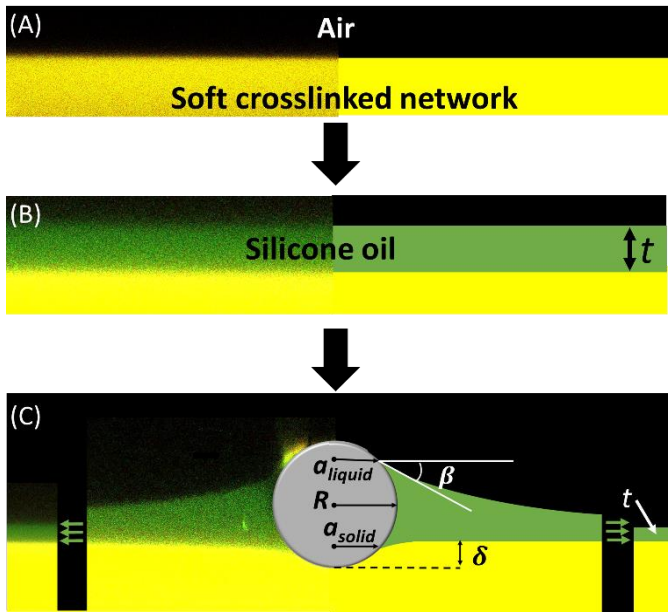

Figure 1. Confocal images (left) are shown alongside schematics (right) of (A) a cured 60 to 1 Sylgard 184 substrate shown in yellow, (B) the substrate with an oil layer shown in green, and (C) the oil-coated soft substrate with a microsphere placed on the surface. In (C), the green arrows denote a distance far from the particle. The schematics provide descriptions of the variables measured from the images.

thickness.[13, 27, 37-40] To investigate the effect of oil layer thickness on the indentation behaviour, we spin coat the uncured Sylgard 184 base (e.g. silicone oil) on top of the cured PDMS (Fig. 1B) with thicknesses ranging from 3 to 40 μm (Fig. S1); this range allows for probing a range of oil layer thicknesses relative to the polydisperse particles. To be able to visualize the PDMS network and the liquid top layer, a Fluorescein fluorescent monomer is incorporated into the crosslinking reaction, and we mix a different Nile Red fluorescent dye into the top oil layer. The modulus of 60 to 1 Sylgard 184 with the fluorescent dye is measured by shear rheology to be $E \approx$ 3.5 ± 0.5 kPa (Fig. S2). This is similar to previously reported moduli, confirming the dye has a negligible effect on the modulus. Additionally, these two dyes have relatively small overlap in their emission wavelengths. Silica microspheres are then sprinkled onto the surface and a cross-sectional image is obtained using confocal microscopy (Figure 1C, left). From the confocal images, we make measurements of the microsphere radius, $R$; the oil contact radius, $a_{liquid}$; the substrate contact radius, $a_{solid}$; the indentation depth into the substrate, $\delta$; the as-coated oil layer thickness, $t$; and the angle of oil contact relative to the horizontal, $\beta$ (Figure 1C, right). By measuring these parameters, we expect to be able to describe and verify the contact behaviour.

When a microsphere is placed on a PDMS surface with a thin oil layer, a liquid meniscus forms and the particle indents into the underlying crosslinked substrate. This is demonstrated in Figure 2A, which shows a ~35 μm diameter glass microsphere in contact with a soft PDMS substrate (yellow) having a ~3 μm oil layer (green). On the other hand, Figure 2B shows a similarly sized particle with an oil layer that is the same thickness as the sphere diameter (e.g. $t \approx 2R$). Unlike in Figure 2A, the sphere does not visibly indent into the underlying crosslinked network. In the other limiting case where no oil layer is present, we find that the particle has a large indentation depth and a large contact area with the network, as illustrated in Figure 2C. When no oil layer is present, the relative indentation depth $\delta/(2R)$ increases as the particle size decreases (Fig. S3). This is consistent with prior results on elastocapillary scale contact showing that indentation is size dependent.[14] In the following, we focus on the indentation as a function of the relative oil layer thickness, $t/(2R)$.

To quantitatively understand how a microparticle indents into an oil-coated surface, we plot the relative indentation depth as a function of the relative oil thickness (Fig. 2D). Additional confocal images of microspheres on surfaces with various oil layer thicknesses are provided in Figure S4. These results show that microspheres indent into the crosslinked network less as the relative oil layer thickness increases. When $t/(2R) \geq 1$, the particle does not indent. Additionally, to test if there is a size dependence on the indentation, we label the particle sizes within a relatively constant $t/(2R)$ range and see if a trend exists in $\delta/(2R)$ (Fig. S5). The lack of an obvious trend between depth and particle size illustrates that the relative oil layer thickness is the dominating factor on the indentation and not the particle size. Therefore, a small amount of oil at the surface transitions the contact from size-dependent (no oil layer) to size-independent (with oil layer).

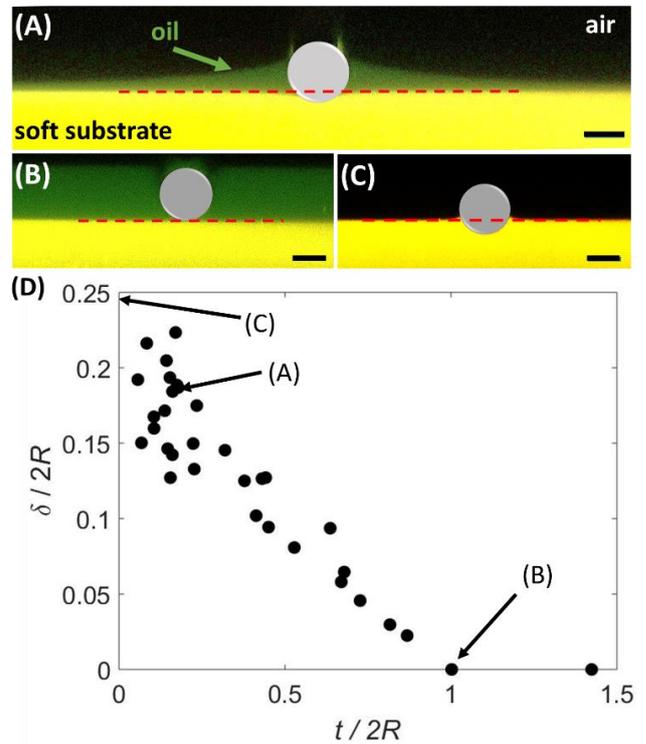

Figure 2. Indentation of a glass microsphere into soft substrates (yellow) with (A) a thin oil layer (green), (B) a thick oil layer, and (C) no oil layer. The dotted line (red) denotes the unaltered substrate surface. The scale bars are 20 μm. (D) Normalized indentation depth as a function of the normalized oil thickness.

The results in Figure 2 illustrate that the oil layer thickness dictates how deep the particle indents into the substrate. This suggests that capillary forces from the oil layer push down on the particle and that the magnitude relates to the oil layer thickness. Since the microsphere is static, the sum of all forces, $F_{total}$, acting on the microsphere must be zero. To describe the indentation, we start by writing out the total force to include the capillary force pushing the microsphere into the surface,[41] the adhesion between the particle and the surface, and the elasticity with the JKR model:

$$F_{Total} = \frac{8Ea_{solid}\delta}{3} - \frac{8Ea_{solid}^3}{9R} - 2\pi\gamma a_{liquid}sin\beta \quad (1)$$

where $E$ is the Young's modulus and $\gamma$ is the liquid oil surface tension. It should be noted that we assume a Poisson's ratio of 0.5 for the PDMS substrate and an infinitely stiff modulus for the glass compared to the PDMS. Moreover, since we are working on small scales, gravity is negligible relative to surface forces. By setting Equation 1 to zero and rearranging for δ, we obtain an expression to predict the particle indentation that includes elasticity and adhesion from JKR balanced by the oil layer capillary force:

$$\delta = \frac{3\pi a_{liquid}\gamma \sin\beta}{4Ea_{solid}} + \frac{a_{solid}^2}{3R} \quad (2)$$

where we take $E = 3.5$ kPa for the PDMS substrate and $\gamma = 20$ mN/m for the silicone oil.[13, 27, 42] Using experimentally measured values for $a_{liquid}$ and $\beta$, we compare δ from Equation 2 to our measured indentation depths (Fig. 3A). We find that Equation 2 predicts a higher indentation depth than experimentally measured; therefore, a non-existing downward force or a missing upward force is not being accounted for.

To consider if the adhesive force in the JKR model is appropriate for our experiments, we investigate the amount of adhesion at the interface. We explore the adhesion using an atomic force microscope (AFM) with a ~20 μm diameter colloidal probe prepared from the same batch of microspheres. In Figure 3B, we display a series of confocal images illustrating a small amount of adhesion between the colloidal probe and the network. In the first image (Fig. 3Bi), the microsphere is held above the oil-coated soft substrate. The microsphere is then pressed into the substrate at a rate of 2 μm/s to a relative depth of ~0.2 (Fig. 3Bii). This indentation depth is chosen to be similar to the recorded indentation depth of free microspheres. The sphere is held for 5 minutes and then retracted at the same rate (Fig. 3Biii and iv). As the sphere is retracted, only a small amount of network pull up is observed, which is indicative of minimal adhesion. We note that upon contact, the oil comes up to contact the cantilever, making it difficult to decouple network adhesion and oil capillarity from measured forces. While this may be due to some extra glue on the particle during colloidal probe fabrication, it does not change the result of finding a small amount of adhesion at the sphere-network interface. We note that this is consistent with the common fitting of Hertzian contact when indenting soft solids in submerged environments.[43, 44]

Since minimal adhesion occurs in the contact, we remove the adhesive component and balance the capillary force against the Hertz model in the total force equation:

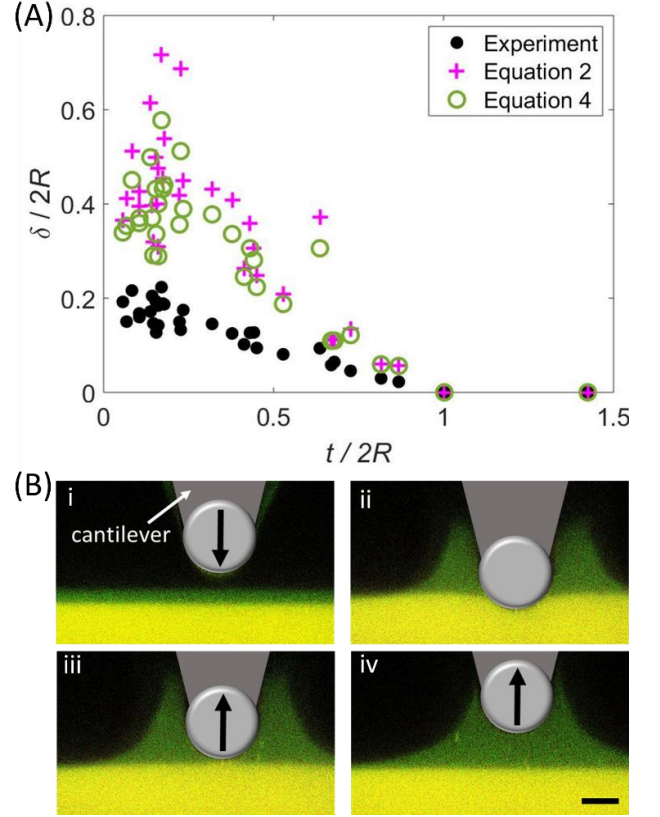

Figure 3. (A) A comparison of the calculated relative indentation depth, $\delta/2R$, from Equation 2 and Equation 4 to the experimental data as a function of the relative oil thickness, $t/2R$. (B) Confocal images of a colloidal probe indentation test when the microsphere was (i) above the oil-coated PDMS before contact, (ii) indented to a relative depth of ~0.2 and held for 5 minutes, and (iii and iv) pulled off the surface at a rate of 2 μm/s. Scale bar is 10 μm.

$$F_{total} = \frac{16}{9}E\sqrt{R\delta^3} - 2\pi\gamma a_{liquid}sin\beta \quad (3)$$

To compare our experiments to Equation 3, we set the total force to zero and solve for δ:

$$\delta = \frac{3}{4}\left(\frac{3}{R}\left(\frac{\pi a_{liquid}\gamma sin\beta}{E}\right)^2\right)^{\frac{1}{3}} \quad (4)$$

Using measured values for $a_{liquid}$, $\beta$, and $a_{solid}$, we compare our measured δ to that predicted by Equation 4 (Fig. 3A). Predicted values from Equation 4 are shifted slightly compared to Equation 2 but are still far from capturing the experimental results. It should be noted that the JKR prediction reduces back to the Hertz prediction when no adhesion is present. Since Equation 2 and Equation 4 are not significantly different, the JKR model is reducing toward the Hertz contact.

It has been previously reported that indentation of microspheres near the elastocapillary scale do not fit JKR due to the importance of solid surface stress.[13, 14, 21, 27, 40, 45] This solid surface tension leads to an additional force that resists deformation during indentation. Therefore, we also consider a solid surface tension term. By calculating the change in the area of a flat plane when indented to form a spherical cap, the force needed to create the additional surface is given as $F_{surface} \approx 2\pi\Upsilon\delta$, which can be incorporated into Equation 1 or Equation 3.[13, 14] Here we first incorporate it into Equation 1 to provide a more universal expression that then includes elasticity and surface stress that resist indentation as well as adhesion and liquid capillary forces that promote indentation:

$$F_{Total} = \frac{8Ea_{solid}\delta}{3} - \frac{8Ea_{solid}^3}{9R} + 2\pi\Upsilon\delta - 2\pi\gamma a_{liquid}\sin\beta \quad (5)$$

This equation is similar to one previously proposed,[13] but separates the contact radius to the solid and liquid ($a_{solid}$ and $a_{liquid}$) since we are able to experimentally visualize these contact lengths experimentally. Equation 5 is then rearranged and solved for a universal indentation depth:

$$\delta = \frac{18\pi a_{liquid}\gamma R\sin\beta + 8Ea_{solid}^3 + 9F_{total}R}{24ERa_{solid} + 18\pi R\Upsilon} \quad (6)$$

Equation 6 can be further simplified by replacing the variable $\beta$ with the liquid contact radius and the sphere radius by using the trigonometric relation:

$$\beta = \sin^{-1}\left(\frac{a_{liquid}}{R}\right) \quad (7)$$

This geometric relation is described schematically in Figure S6 and allows us to use the more easily measurable $a_{liquid}$ instead of the horizontal angle $\beta$. Additionally, in our experiments the total net force is zero, which yields:

$$\delta = \frac{9\pi\gamma a_{liquid}^2 + 4Ea_{solid}^3}{12ERa_{solid} + 9\pi R\Upsilon} \quad (8)$$

The indentation depth predicted by Equation 8 is compared to the measured indentation depth by using measured contact geometries (Fig. 4A). Here we approximate the solid surface tension to be the same as the liquid tension, $\Upsilon = 20$ mN/m.[27] It was recently shown in a numerical study that the solid surface tension of a soft solid and a polymer melt are similar until high strains are reached.[46] We do not expect the strains to be large enough to significantly modify $\Upsilon$. The predicted values overlay closely to the measured indentation depth without any fitting parameters. However, this equation includes an adhesive component that did not significantly change the predicted indentation depth when comparing Equations 2 and 4 (Fig. 3A); therefore, the adhesive component of Equation 5 may be able to be removed in our specific case. By balancing liquid capillary force with the Hertz model and surface stress, we come to a total force equation specific to the case of no network adhesion:

$$F_{total} = \frac{16}{9R}Ea_{solid}^3 + 2\pi\Upsilon\delta - \frac{2\pi\gamma a_{liquid}^2}{R} \quad (9)$$

Here we assume that the depth follows the Hertz relation $\delta = a_{solid}^2/R$ to simplify the algebraic expression.[47] Solving this equation for the indentation depth yields:

$$\delta = \frac{9\pi a_{liquid}^2\gamma - 8Ea_{solid}^3}{9\pi R\Upsilon} \quad (10)$$

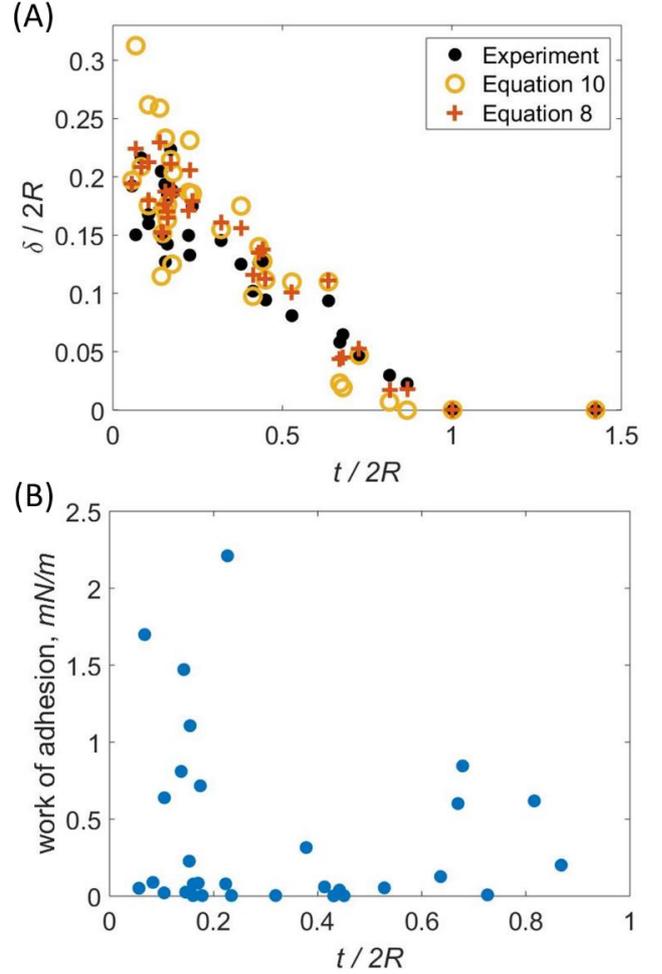

Figure 4. (A) The relative indentation depth predicted by Equation 8 and Equation 10 overlaid on experimental data. (B) The work of adhesion calculated by the rearranged form of Equation 5.

Equation 10 also shows a reasonable overlay of the measured data without any fitting parameters (Fig. 4A). Interestingly, this reveals that a Hertzian type contact can also require surface stress when adhesion is not a dominating factor.

By looking at the predicted values from Equation 10 (modified Hertz), we observed more deviation from the experimental measurements than with Equation 8 (modified JKR). To investigate the possible reason, we considered if the work of adhesion $w$ from Equation 5 is actually zero. Equation 5 is rewritten to include the work of adhesion term as $F =$

$16Ea_{solid}^3/(9R) - (32\pi a_{solid}^3 Ew/3)^{1/2} + 2\pi Y\delta - 2\pi\gamma a_{liquid}\sin\beta$ and then solved for $w$ with measured contact geometry (Fig. 4B). Although the majority of the calculated $w$ are zero, some have values of up to ~3 mN/m. This is indeed small but nonzero, and these data points are the ones that deviate more from our experimental measurements of indentation depth. These discrepancies may arise from inhomogeneities at the contacting interface, pinning effects, or the resolution of our measurements. The results in Figure 4 illustrate that Equation 6 is more universal for capturing the indentation depth and Equation 10 is valid only when the apparent work of adhesion is zero.

## Conclusion

In summary, we have shown that the presence of a thin oil layer leads to the formation of an oil meniscus around a microsphere, which relates to a downward capillary force. This suggests that the addition of an oil layer transitions the balance of forces from solid adhesion dominated to liquid capillary dominated. We find that the downward capillary force reduces as the thickness of the oil layer increases. A model that includes elasticity, adhesion, surface stress, and liquid capillary forces is able to capture the experimental results. Moreover, when a thin oil layer is present, solid adhesion is minimized and a modified Hertz model that includes surface stress can be balanced against the capillary forces of the oil. Understanding small scale contact on a soft oil-coated surface will be beneficial for bioinspired adhesives,[8, 48, 49] soft tribology,[12, 31, 50-60] soft robotics,[61, 62] and anti-fouling self-cleaning coatings.[63-66]

## Experimental

***Materials.*** Dow Sylgard 184 was purchased as a two-part kit from Ellsworth Adhesives. Polydisperse soda lime glass microspheres were purchased from Cospheric LLC. The particles in this study ranged in size from ~9 μm to 31 μm in radius. 22x30 mm, No. 1 glass coverslips and chloroform were purchased from VWR. Nile red was purchased from Acros Organics. Fluorescein diacrylate was purchased from Sigma-Aldrich.

***PDMS (Polydimethylsiloxane) preparation.*** Sylgard 184 base, which is comprised of vinyl-terminated polydimethylsiloxane, is mixed with the curing agent, comprised of methylhydrosiloxane– dimethylsiloxane copolymer and a catalyst.[48] The two parts were mixed in a ratio of 60 to 1 base to crosslinker and degassed under vacuum to remove any trapped air, ~30 minutes. The solution was spin coated on a glass coverslip at 800 RPM for 60 seconds to achieve a thickness of ~90 μm. Other RPMs can be used to increase or decrease the thickness (Fig. S1). An RPM of 800 was chosen to maximize the thickness of the PDMS while maintaining the resolution using an optically correctable objective. The coverslip with the uncured PDMS is cured in an oven at 65 °C for 48 hours.

*Fluorescein diacrylate addition.* ~0.005 g of fluorescein diacrylate was dissolved in a minimal amount of chloroform (~1 mL) and added to ~7 g of Sylgard 184 base. The concentration of the fluorescein diacrylate in the base was approximately ~0.5 mg/g. Next, the solution was placed in an oven at 65 °C to evaporate the added chloroform. After 4 days the weight of the solution stabilizes showing that all the chloroform is removed (Fig. S7). Then, the base with the fluorescein diacrylate was used in the PDMS preparation processes described above. Fluorescein diacrylate was chosen as the dye for the substrate because it is expected to react with the vinyl-terminated ends of the prepolymer base.

*Silicone oil layer.* Nile red was dissolved in chloroform and added to Sylgard 184 base in the concentration of approximately 5 μg Nile Red/1 g Sylgard 184 base. The solution was heated in an oven at 65 °C until all the chloroform was evaporated. To form the oil layers on the surface of the PDMS, the Sylgard 184 base with Nile red was spin coated on the surface of cured PDMS at various RPMs and durations. As a baseline, 6000 RPM for 120 seconds produced an oil layer of approximately 8 microns (Fig. S1).

***Characterization.*** *Modulus.* 60 to 1 Sylgard 184 dyed with Fluorescein diacrylate was prepared following the previously described procedure but was cured in a 35 mm diameter Petri dish to form ~1 mm thick samples. Four samples were made from 2 different batches of 60 to 1 Sylgard 184. The samples were tested using a TA Instruments Discovery HR-2 rheometer using 25 mm parallel plates. The storage modulus of each sample was tested to a strain of 0.5% at a rate of 0.01 rad/s after confirming this strain was in the linear region of a strain sweep for each sample (Fig. S2). The Young's modulus was then calculated from the shear storage modulus by assuming a Poisson's ratio of 0.5.

*Imaging via confocal microscopy.* Individual microspheres, sprinkled on the samples of oil-coated surfaces, were imaged using a Leica SP8 inverted confocal microscope with a piezo driven 40x air objective. Once oil was spin coated onto a sample and particles sprinkled, the sample was left to equilibrate for 30 minutes and imaged within 1 hour of spin coating the oil. Microspheres were selected that were ~1 mm from another microsphere to avoid affects from other microspheres.

*Image analysis.* The confocal images were analyzed using ImageJ. A sphere was fit to the shape of the particle in the image, and the distance from the lowest point of the sphere to the top of the PDMS network was recorded as the indentation depth. For samples containing an oil layer, the height of the oil was determined by measuring the top of the network to the top of the oil outside of the meniscus of the oil caused by the particle.

*Colloidal probe microscopy.* A JPK Nanowizard 4 was used to perform AFM the adhesion tests. A ~20 μm diameter glass sphere from the microspheres used in the free particle test was attached to a tipless cantilever with a 31.7 N/m stiffness using high strength epoxy. The indentation and pull-off rates were 2 μm/s. The particle was pressed into the substrate to a relative indentation depth of ~0.2 and held for 5 minutes before pull-off.

## Acknowledgements

We are grateful for financial support by the National Science Foundation (CMMI-1825258) and University of Kentucky start-up funds.